\documentclass[12pt]{article}

\usepackage{scicite}

\usepackage{times}

\usepackage{graphicx}

\usepackage{amsmath}

\usepackage{caption}

\usepackage{siunitx}

\newenvironment{sciabstract}{%
\begin{quote} \bf}
{\end{quote}}

\title{Inverse-designed low-index-contrast structures on silicon photonics platform for vector-matrix multiplication}

\author
{Vahid Nikkhah,$^{1}$ Ali Pirmoradi,$^{1}$ Farshid Ashtiani,$^{1}$\\
Brian Edwards,$^{1}$ Firooz Aflatouni,$^{1}$ Nader Engheta$^{1\ast}$\\
\\
\normalsize{$^{1}$Department of Electrical and Systems Engineering, University of Pennsylvania,}\\
\normalsize{Philadelphia, PA 19104, U.S.A.}\\
\\
\normalsize{$^\ast$To whom correspondence should be addressed; E-mail:  engheta@seas.upenn.edu}
}

\date{}

\begin{document} 


\baselineskip24pt


\maketitle


\begin{sciabstract}
    Inverse-designed Silicon photonic metastructures offer an efficient platform to perform analog computations with electromagnetic waves.
    However, due to computational difficulties, scaling up these metastructures to handle a large number of data channels is not trivial.
    Furthermore, a typical inverse-design procedure utilizes a small computational domain and therefore tends to employ resonant features to achieve its objectives.
    This results in structures that are narrow-bandwidth and highly sensitive to fabrication errors.
    Here, we employ a 2D inverse-design method based on the effective index approximation with a low-index contrast constraint.
    This results in compact amorphous lens systems which are generally feed-forward and low-resonance.
    We designed and experimentally demonstrated a vector-matrix product for a $2 \times 2$ and a $3 \times 3$ matrix.
    We also designed a $10 \times 10$ matrix using the proposed 2D computational method.
    These examples demonstrate that these techniques have the potential to enable larger-scale wave-based analog computing platforms.
\end{sciabstract}


\section*{Introduction}
Photonic structures that can perform mathematical operations and solve equations are becoming increasingly popular due to the resurgence of optical analog computing with the promise of low-power, high-speed, parallel computations enabled by light\cite{cordaro2022solving,nikkhah2022inverse,silva2014performing,mohammadi2019inverse,zhang2021optical,camacho2021single,fu2022ultracompact,pan2021laplace,zangeneh2021analogue,zhu2017plasmonic,pors2015analog}.
Specifically, neural networks can benefit from vector-matrix multiplication rendered as optical analog computing modules with improved energy consumption per arithmetic operation while also achieving significantly higher speeds\cite{wang2022optical,shastri2021photonics,shen2017deep,lin2018all,feldmann2021parallel}.
Physically defining a $N\times N$ matrix for vector-matrix multiplication requires the realization of $N^2$ different objectives within a single device.
Performing vector-matrix multiplication with electromagnetic waves can be formulated either as free space Fourier optics in k-space\cite{silva2014performing,wang2020compact,zhu2021topological,cordaro2022solving} or as spatial modes as the basis for performing in real space\cite{mohammadi2019inverse,goh2022nonlocal}.
This can be done in one of two ways.
The $n$ modes can be expanded into $n^2$ modes, utilizing an additional dimension. These are directly acted upon and then summed\cite{farhat1992photonic}. 
Within a 3D device, such an operation can be achieved using metasurfaces defined under the Born approximation considering only their transmission coefficients\cite{wetzstein2020inference,farhat1992photonic}.
Alternatively, the $n$ modes can be operated through a series of mixing operations in the form of a mesh architecture, generally "forward" only\cite{miller2013self,bogaerts2020programmable,clements2016optimal,tzarouchis2022mathematical}.
Finally, the modes may be operated on by an inverse-designed structure, potentially containing internal resonances\cite{mohammadi2019inverse,cordaro2022solving}.

Using a combination of topological and shape optimizations, Inverse Design\cite{mohammadi2019inverse,cordaro2022solving} is particularly well suited for designing such structures. 
Specifically, density-based topology optimization is a technique in which the material within a design region is suitably discretized into a large number of free parameters without any preconceived idea of the nature of the structure\cite{bendsoe2003topology}.
These parameters are then optimized utilizing a suitable mathematical technique such as the adjoint method which can compute the gradient utilizing a comparatively few numbers of simple forward simulations\cite{molesky2018inverse,hughes2018adjoint}. 
Following topological optimization, the material can then be binarized and vectorized into highly parameterized shapes which are similarly optimized using the adjoint method. 

Although a structure in silicon photonics (SiPh) platform is planar in nature, the fields do not possess 2D symmetry, forcing designers to often utilize 3D simulations\cite{hammer2009effective}.
Due to the computational difficulties of 3D simulations and the potentially large number of iterations needed to reach optimal performance, inverse design in 3D is limited to optically small structures (i.e. tens of cubic wavelengths).
This effectively limits the number of information channels and therefore the matrix size that may be considered.
There is a need for an efficient computational platform that reduces the computational costs of full-wave 3D simulations in SiPh.

Therefore, a "reduced order" approximation for photonic structures that can perform analog computation between a large number of input and output information channels is vital.
In this paper, we employ a propagation-based 2D effective index approximation (p2DEIA)\cite{knox1970integrated,chiang1986analysis,hocker1977mode,van1988extending} for 3D planar structures which can significantly reduce the computational effort.
By designing SiPh structures with relatively small variations in Silicon thickness, the effective index difference between these regions will be small.
This \emph{low-index contrast} reduces out-of-plane scattering, increases the accuracy of the p2DEIA approximation, and reduces reflections (i.e. reduces resonances and therefore increases bandwidth).
We validate the inverse design method based on the low-index-contrast p2DEIA by designing and fabricating SiPh metastructures that perform vector-matrix multiplication.

\section*{Results}
\subsection*{Problem Description}
\begin{figure}[!ht] 
    \centering
    \includegraphics[width = \textwidth]{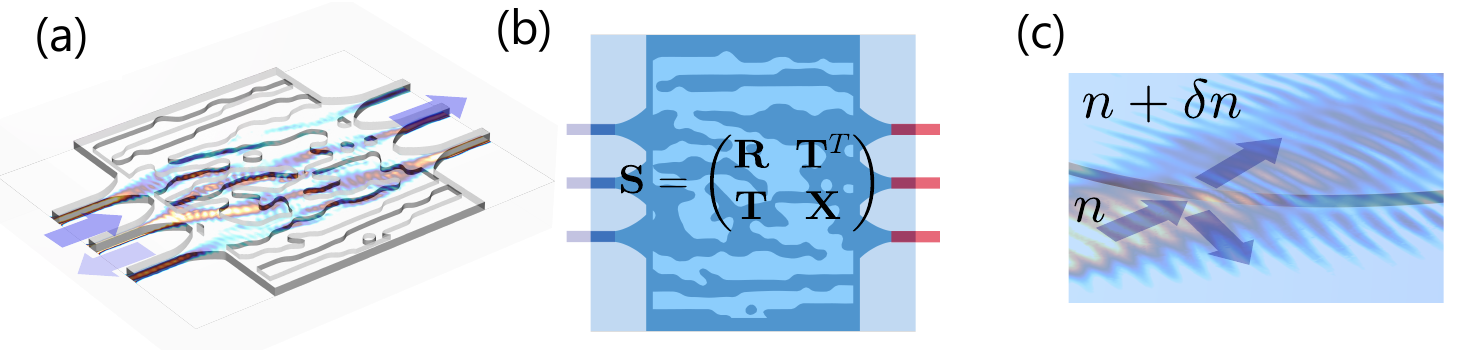}
    \caption{\textbf{Schematic of our inverse-designed structures in Si-photonic platform} (a) Vector-matrix multiplication consisting of transmissive part of the scattering matrix, $S$.(b) Generic structure which is optimized to achieve this transmissive part. (c) Feed-forward due to low-index contrast leading to small reflection and primarily refraction behavior.}
    \label{Concept Figure}
\end{figure}
In this paper, we perform vector-matrix multiplication for a given matrix $\mathbf{M}$ on a SiPh platform using the complex amplitude of a series of spatial modes to represent our vectors.
For that purpose, an array of single-mode silicon waveguides is considered as the input channels and a second array represents the output channels.
Both the input and output waveguides are connected to a design region in which the Silicon thickness is parameterized, varying between two values. (see Fig.\ref{Concept Figure}(a)).
In general, such a structure can be fully described using a scattering matrix, in which the incoming spatial modes on any waveguide (designated input or output) will map to all others (see Fig.\ref{Concept Figure}(b)).
The waves injected through the designated input ports can be mathematically described as a complex-valued vector. 
Despite being secondary to the design, an incoming vector from the designated output ports can be similarly defined.
By representing the incoming waves as two concatenated $N\times 1$ vectors, the scattering matrix ($2N \times 2N$) can be defined using four block matrices (each $N \times N$).
Since we are interested in the propagation from the input to the output ports as the physical process for realizing a vector-matrix product, we are primarily concerned with the transmission block matrix, ($\mathbf{T} = \mathbf{M}$). 
Additionally, we would like to minimize the reflections back towards the designated input ports (i.e. $\mathbf{R} = \mathbf{0}$).
Our goal, therefore, is to compute a spatially varying Silicon thickness distribution within the 2D design region, which achieves these objectives.

\subsection*{Effective-Index Approximation}
It is known that the simulation and design of a 3D structure is a computationally costly process, especially in any iterative design method.
The confined wave within the silicon slab propagates through the in-plane geometry in the form of a guided slab mode. 
The effective index of a slab mode depends on the thickness of the slab and the refractive index of the substrate, silicon core, and cladding. 
For example, a $220\ nm$-thick silicon layer with $n_\text{Si} = 3.48$ immersed in $\text{SiO}_2$ with $n_{\text{SiO}_2} = 1.44$ supports a fundamental slab mode with $n_\text{eff, 220nm} = 2.86$ at $\lambda_0 = 1.525 \ \mu m$.
The effective index decreases as the thickness of the silicon core decreases.
For example, for a similarly defined $150 \ nm$-thick silicon layer, the effective index of the fundamental slab mode reduces to $n_\text{eff, 150nm} = 2.56$. 
Within certain constraints, one can construct a 2D computational model which approximates the in-plane wave propagation (i.e. the guided slab modes) through a 3D planar structure by using the effective indices of the slab modes to represent the various silicon core thicknesses.
For instance, a region in the 2D model with a refractive index of $n = n_\text{eff, 220nm} = 2.86$ fully and accurately models the in-plane propagation of the fundamental slab mode in the 3D structure with a silicon thickness of $220 \ nm$.
This model is here referred to as the propagation-based 2D Effective Index Approximation (p2DEIA).

Suppose we have two regions, one with a Silicon layer thickness of 220nm, and the other which has been uniformly etched to achieve a smaller thickness (see Fig.\ref{Concept Figure}(c)).
When the fundamental slab mode hits the interface between these two regions, various phenomena can happen.
The p2DEIA model will effectively capture the angle of the reflection and transmission of the fundamental slab modes through Snell's law (see Fig.\ref{Concept Figure}(c)).
However, a 3D height discontinuity will experience other effects.
For instance, some portion of the wave is scattered away from the silicon slab into the $\text{SiO}_2$ cladding.
Additionally, the excitation of higher-order slab modes inside the silicon core may occur. 
With these additional scattering channels, a 3D height discontinuity may \emph{only} be approximated by a 2D model.
The accuracy of the 2D model depends on the strength of the coupling to these other scattering channels. 
When the silicon thicknesses are similar, the modal overlap between the fundamental slab modes is very large, and hence, all other couplings must be small (see SM Figure 1)
Therefore, for shallow-enough etch depths, one can assume the coupling to the unmodelled open channels is negligible and safely use the approximate p2DEIA method with good accuracy, even at interfaces.

The p2DEIA method based on the index of the fundamental slab mode defined previously accurately captures the propagation within a uniformly etched region, angle of refraction, and angle of reflection.
However, it may not capture the magnitude and phase of the reflection and transmission at an interface well compared to other models.
For instance, in models based on variational methods, one chooses a thickness where one expects most of the energy of the wave to reside\cite{hammer2009effective}.
The effective index of this region is calculated similarly to the above and accurately models propagation.
However, all other effective indices are defined in reference to this chosen region and are designed to accurately model the interface behavior, at the expense of modeling propagation in these other regions accurately\cite{hammer2009effective}.
These variational methods may be preferred for waveguiding structures based on total internal reflection (TIR) and grating couplers, where the accurate modeling of the interfaces between regions is crucial.  
However, they will be less accurate when significant energy is propagating across etched regions for extended distances, where phase error due to compromised affected indices will accumulate.
We will refer to this as the interface-based effective index approximation (i2DEIA).

As mentioned, the accuracy of the p2DEIA requires that the relative difference in silicon thickness across an interface is not large, and we will limit ourselves to designs in which this statement is true.
As a result, the effective indices associated with each region will have \emph{low-index contrast}, rendering the interface reflection magnitude to be small and transmission magnitude to be large.
Therefore, one can expect that the wave mostly travels forward from input to output ports, diffracting at interfaces as it moves from region to region.
The lack of significant reflections will result in a low-resonance feed-forward optical system that behaves similar to a series of irregular-shaped lenses.
This type of system will naturally present broadband optical responses and will be less sensitive to small perturbations in the shape of the interfaces so long as the angles of the interfaces are generally maintained.
These are useful features from an experimental point of view in which a metastructure that is robust against fabrication errors such as "over etching" is highly desired.
On the other hand, the low-index contrast interfaces will provide only weak diffraction and will offer only weak control over wave manipulation compared to the highly-contrasted interfaces.

\subsection*{Effects of Passivity on Geometry Definition}
Any photonic structure that is enclosed by a boundary $S$ can be described as a unitary scattering matrix, provided that it exhibits only lossless propagation.  
The modes on such a boundary can be enumerated.  
Some of these modes will be designated input ($\mathbf{I}$), output ($\mathbf{O}$) and the remainder will be considered simply ``absorbing'' ($\mathbf{A}$).
In all cases, any outgoing waves on these modes ($\mathbf{I}_\mathrm{out}$, $\mathbf{O}_\mathrm{out}$, $\mathbf{A}_\mathrm{out}$) will not return to the system.
In addition to block matrices $\mathbf{T}$ and $\mathbf{R}$, the addition of the vector $\mathbf{A}$ allows us to conceptualize other block matrices as well ($\mathbf{\chi}_1$, etc).
\begin{equation}
\begin{pmatrix}
\mathbf{I_\mathrm{out}}\\
\mathbf{O_\mathrm{out}}\\
\mathbf{A_\mathrm{out}}\\
\end{pmatrix} = 
\mathbf{S} \begin{pmatrix}
\mathbf{I_\mathrm{in}}\\
\mathbf{O_\mathrm{in}}\\
\mathbf{A_\mathrm{in}}\\
\end{pmatrix} 
\hspace{1in}
\mathbf{S} = 
\begin{pmatrix}
\mathbf{R} & \mathbf{T}^T & \mathbf{\chi_1}\\
\mathbf{T} & \mathbf{X} & \mathbf{\chi_2}\\
\mathbf{\chi_3} & \mathbf{\chi_4} & \mathbf{\chi_5}\\
\end{pmatrix}
\end{equation}
These absorbing modes will never be illuminated (i.e. $\mathbf{A}_\mathrm{in} = \mathbf{0}$) and the precise shape of these modes will not be characterized.
However, these modes do affect the system.

Since the scattering matrix is unitary, all singular values of $\mathbf{S}$ are one.
As a block matrix within a unitary matrix, the singular values of $\mathbf{T}$ must be less than one.
This condition is known as passivity.
When the number of unconstrained absorbing modes is very large, then $\mathbf{S}$ being unitary ceases to be a significant constraint, and the condition of passivity on $\mathbf{T}$ is sufficient.
On the other hand, if the number of absorbing modes is small (i.e., one), there is no guarantee that even a scattering matrix defined with a passive $\mathbf{T}$, $\mathbf{R} = \mathbf{0}$ and unconstrained $\mathbf{X}$ belongs to the space of unitary matrices.
Therefore, within this work, we assume that the unconstrained $\mathbf{A}$ represents a sufficiently large number of absorbing modes.
In that case, we may use the passivity condition to select the target transmission matrices.
Within the vector space described by the passivity constraints on the transmission matrix, we arbitrarily choose the following matrices for a $2\times 2$ and $3 \times 3$ optical network:
\begin{equation}
\mathbf{T}_\mathrm{2\times2} = 
\begin{pmatrix}
0.43+j0.43 & -0.47+j0.12 \\
0.43+j0.22 & 0.51-j0.33\\
\end{pmatrix}
\hspace{1in}
\mathbf{R}_\mathrm{2\times2} = \mathbf{0}
\end{equation}

\begin{equation}
\mathbf{T}_\mathrm{3\times3} = 
\begin{pmatrix}
0.49+j0.24 & -0.53 + j0.44 & 0.44 - j0.16\\
0.60+j0.41 & 0.20-j0.46 & -0.32-j0.36\\
0.41+j0.11 & 0.5+j0.15 & 0.20+j0.71\\
\end{pmatrix}
\hspace{1in}
\mathbf{R}_\mathrm{3\times3} = \mathbf{0}
\end{equation}

\subsection*{Metastructure Configuration and Inverse Design}
\begin{figure}[!ht]
    \centering
    \includegraphics[width = \textwidth]{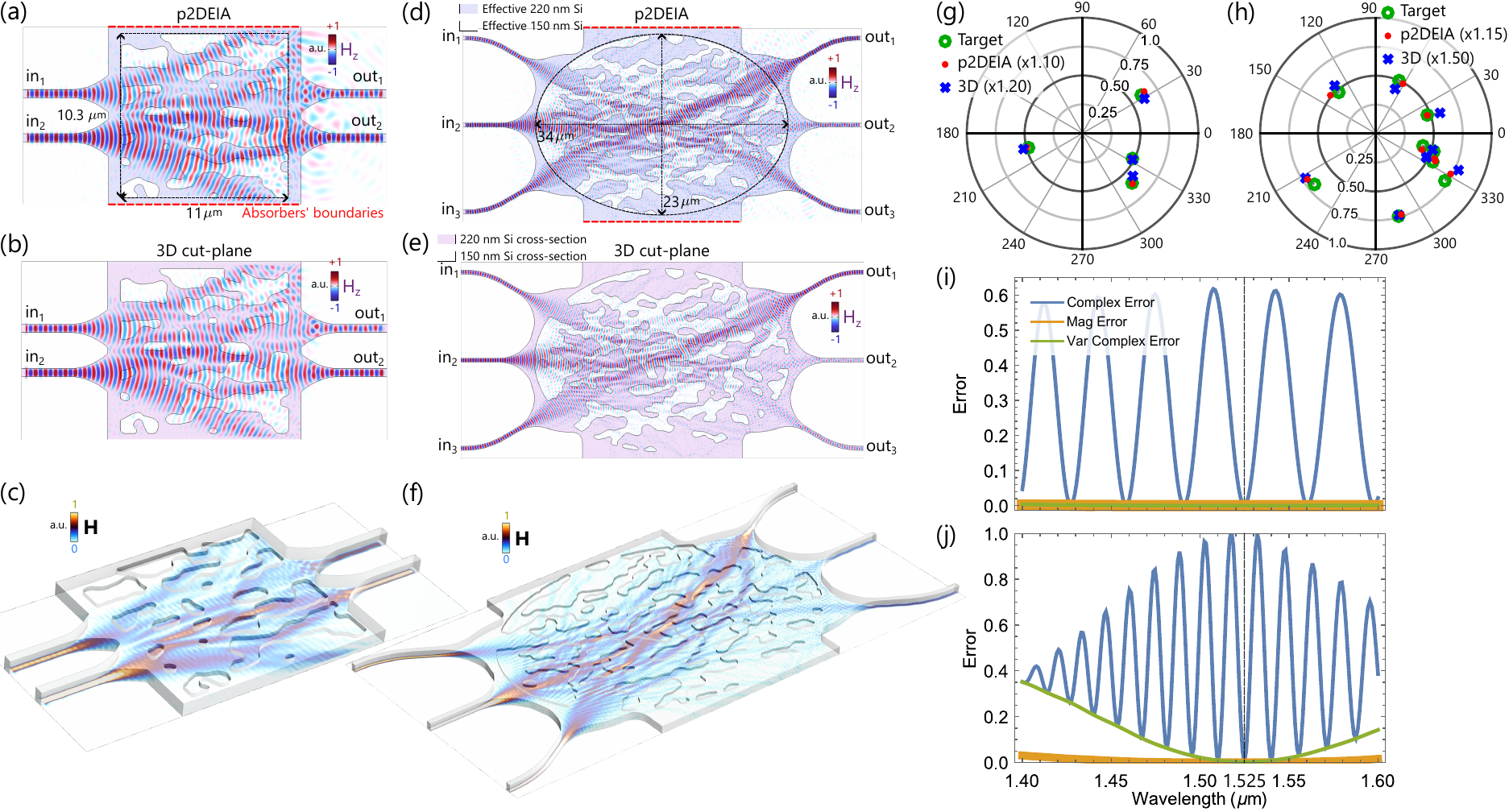}
    \caption{\textbf{Inverse-designed metastructures based on silicon photonics for performing $2 \times 2$ and $3 \times 3$ vector-matrix multiplications.} The designs and simulations are carried out at free-space wavelengths of $\lambda_0 = 1.525 \ \mu m$. The optimized structures in p2DEIA and time snapshots of out-of-plane magnetic field distributions for arbitrary input vector excitations for (a) $2 \times 2$ and (d) $3 \times 3$ examples. (b) and (e) Time snapshot of the magnetic field distributions on cut-planes through silicon in the full-wave 3D-rendered structures for the same excitations as in p2DEIA.  Here we choose a vertical view for easy comparison to the results in the p2DEIA case, i.e. (a) and (d).  (c) and (f) The absolute value of magnetic field distributions through the silicon in the 3D view of the designed structures. (g) and (h) The target transmission matrix element (green circles), the p2DEIA values (after proper adjustment factors shown in parentheses) (red circles), and the 3D values (after proper adjustment factors shown in parentheses) (blue circles) for $2 \times 2$ and $3 \times 3$ structures. (i) and (j) The normalized Frobenius norm of the error squared (i.e. $\lVert \mathbf{T} - \mathbf{T}_\text{targ} \rVert_F^2/4N$) between target and 3D transmission values vs. wavelength for $2 \times 2$ and $3 \times 3$.}
    \label{Des_Sim_Results}
\end{figure}

The proposed optical metastructures for realizing the target transmission matrices are designed based on a silicon photonics platform at $\lambda_0 = 1.525 \ \mu m$. 
The 2D schematics of the optimized $3\times 3$ and $2\times 2$ structures using the p2DEIA model are shown in Figure \ref{Des_Sim_Results}(a) and (d) respectively.
The width of each silicon waveguide is $500 \ nm$ with a thickness of $220 \ nm$. 
For the $2 \times 2$ case, the design region, which is distinguished by the domain inside the dashed black boundary in Figure \ref{Des_Sim_Results}(a), is a rectangular domain of width $11 \ \mu m$ and length $10.3 \ \mu m$.
For the $3 \times 3$ case, it is an ellipse with a width of $34 \ \mu m$ and length $23 \ \mu m$ (see in Figure \ref{Des_Sim_Results}(d)).
In both structures, the absorbers' boundaries, illustrated by dashed red lines, span many wavelengths across the sides, providing a large enough number of absorbing modes $\mathbf{A}$ required as part of the design process.

The optimization problem is mathematically described by the following 
\begin{equation} \label{obj}
    \begin{split}
        & \min_{n(x,y)} \lVert \mathbf{T} - \mathbf{T}_\text{targ} \rVert_F + \lVert \mathbf{R} - \mathbf{0} \rVert_F \\
        & \text{s.t.} \ \ n_\text{eff, 150nm} \leq n(x,y) \leq n_\text{eff, 220nm} \\
    \end{split}
\end{equation}
where $\lVert \cdot \rVert_F$ is the Frobenius norm.
The objective scalar defined for inverse design is the "distance" between the complex-valued target transmission matrix and the transmission matrix that is realized by the current structure in each iteration.
A similar expression for minimizing reflections is also added to the objective scalar as well.
Using p2DEIA the design region is parameterized by an effective index distribution which is bound between $n_\text{eff, 150nm}$ and $n_\text{eff, 220nm}$.
By employing the inverse design method in COMSOL Multiphysics\textsuperscript{\textregistered}\cite{COMSOL} using the density-based topology optimization based on the adjoint method, the effective index distribution within the p2DEIA model is optimized to minimize the scalar objective.

During the inverse design process, the effective index distribution is encouraged to the extremum values of $n_\text{eff, 150nm}$ and $n_\text{eff, 220nm}$ using a sigmoidal projection function which becomes successively steeper.
Then, the optimized distribution is vectorized to define two classes of domains associated with $n_\text{eff, 220nm}$ and $n_\text{eff, 150nm}$, shown as the highlighted and the complementary unhighlighted regions, respectively, within the design regions in Figure \ref{Des_Sim_Results}(a) and (d).
These domains ultimately define corresponding thicknesses in the 3D structure both for simulation and fabrication as shown in Figure \ref{Des_Sim_Results}(c) and (f).

The p2DEIA model is a computationally efficient approximation and is judged based on how well it compares to full 3D simulations of the structure, named here as the ground truth.
Figure \ref{Des_Sim_Results}(a) and (d) show time snapshots of out-of-plane magnetic field distribution in the vectorized p2DEIA model while panels (b) and (e) show the corresponding field distribution on a cut-plane through the silicon slab in the 3D-rendered structure for the same excitation as in the p2DEIA model.
Visually comparing the simulated field distributions of the p2DEIA model to the 3D structure shows similar features, indicating the success of the p2DEIA in approximating the in-plane wave propagation through the 3D structure.
Finally, Figure \ref{Des_Sim_Results}(c) and (f) illustrate the absolute value of the magnetic field distributions on cut-planes through the silicon in the 3D view of the designed metastructures.

In Figure \ref{Des_Sim_Results}(g) and (h) we compare the complex values for the target transmission matrix ($\mathbf{T}_\mathrm{targ}$), that realized for the p2DEIA model ($\mathbf{T}_\mathrm{p2DEIA}$), and for the 3D model ($\mathbf{T}_\mathrm{3D}$).
There are two classes of errors that we allow ourselves to normalize out. 
First, a constant magnitude adjustment is applied to all values in $\mathbf{T}_\mathrm{p2DEIA}$ to account for the unmodeled energy loss due to out-of-plane scattering.
Second, a constant phase adjustment for each class of waveguide is applied to all values in $\mathbf{T}_\mathrm{p2DEIA}$.
This is due to the ineffectiveness of the p2DEIA in modeling waveguides compared to the i2DEIA model.  
In other words, the guided modes in waveguides present a slightly different wavenumber which introduces a phase error between p2DEIA and both the i2DEIA and 3D models. 
Once these trivial adjustments are made, we can see the close agreement between the target, p2DEIA, and the 3D values, indicating the success in achieving the desired transmission coefficients and the good accuracy of the p2DEIA in approximating 3D structures.

Next, in panels (i) and (j) we examine the spectral response of the realized 3D design by examining three types of errors.
In all cases, we normalize the squared error by $4N$, which renders a unitary matrix of size $N$ with 180deg phase error to 100\%.  See SM.
The orange curve is the Frobenius distance between the absolute values of the transmission parameters.  ($\lVert |\mathbf{T}_\mathrm{3D}(\lambda)| -  |\mathbf{T}_\mathrm{targ}| \rVert_F^2/4N$).
The small distance between the target values and the simulated ones indicates that the magnitude of the transmission parameters remains fairly invariant and close to the target as the wavelength changes.
However, as a device that is intended to perform \emph{complex} matrix multiplication, this is an inadequate indicator of success.
The blue curve is the Frobenius distance between the transmission parameters over a range of wavelengths and the target values $\lVert \mathbf{T}_\mathrm{3D}(\lambda) - \mathbf{T}_\mathrm{targ} \rVert_F^2/4N$. 
This distance is quite small at the design wavelength indicating the high accuracy of the 3D structure in realizing the target transmission parameters once the aforementioned amplitude and phase corrections were applied.
However, the error presents an oscillatory behavior as the wavelength changes. 
This is due to the propagation delay between the input and output ports which introduces a global phase error on the transmission parameters.
This is conceptually related to the rotation of phase that occurs upon changing wavelength when light propagates in the forward direction through any medium (i.e. $e^{j (k(\lambda) - k(\lambda_0)) l}$).
This global phase is immaterial as far as the phase differences are concerned, and these phase differences are typically what matters.
Therefore, we consider an error in which we apply a global phase correction (green curve), $\phi(\lambda)$, to all values such that  $\mathbf{T}_\mathrm{3D}^\prime(\lambda) = \mathbf{T}_\mathrm{3D}(\lambda)e^{-j \phi(\lambda)}$ which minimizes $\lVert \mathbf{T}^\prime_\mathrm{3D}(\lambda) - \mathbf{T}_\mathrm{targ} \rVert_F^2/4N$.
As expected, this removes the oscillatory nature of the error and we may conclude that the designed metastructure is capable of performing vector-matrix operations over an extended range of wavelengths, hence being a broadband optical structure for such analog computation.

\subsection*{Experimental Results}
\begin{figure}[!ht]
    \centering
    \includegraphics[width = \textwidth]{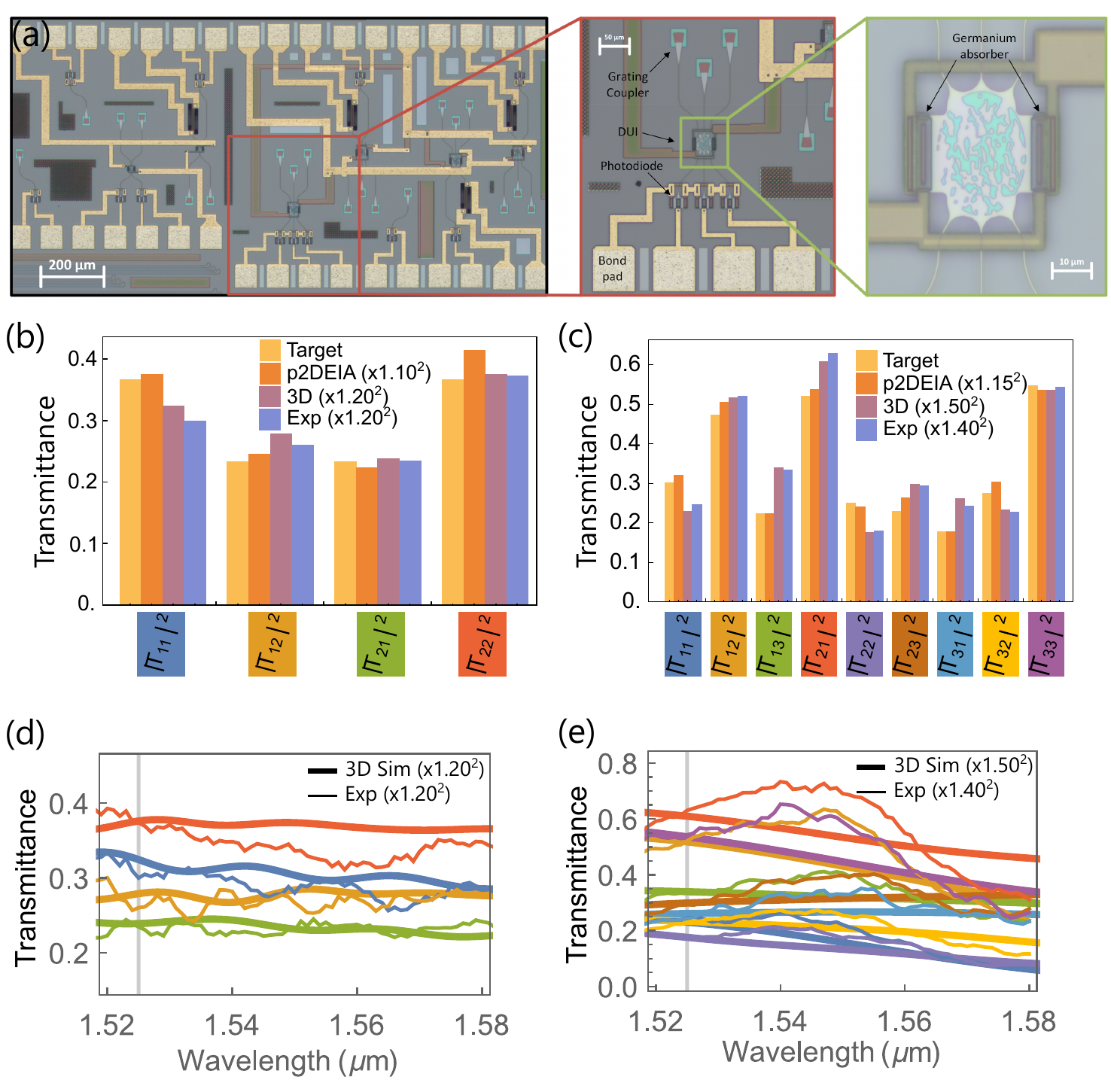}
    \caption{ Comparison between simulations and experimental results.  
    (a) Micrograph of the experiment showing several $2 \times 2$ and $3 \times 3$ kernels and various calibration structures.  The $3 \times 3$ kernel is highlighted.
    (b) and (c) The measured transmittance values are compared with the simulated transmittance values in p2DEIA and 3D models and the target values at $\lambda_0 = 1.525 \ \mu m$ for $2 \times 2$ and $3 \times 3$ metastructures, respectively. 
    (d) and (e) The measured transmittance values (thick solid lines) and the simulation results in 3D (thin solid lines) vs. the wavelength for $2 \times 2$ and $3 \times 3$ metastructures, respectively.}
    \label{experiment}
\end{figure}

The designed metastructures are fabricated and the results of the experiment are illustrated in Figure \ref{experiment}.
In Figure \ref{experiment}(a) and (c), the target transmittance parameters, the simulated transmittance in p2DEIA, the simulated values in the 3D model, and the measured transmittance at $\lambda_0 = 1.525 \ \mu m$ are compared with each other for the $2 \times 2$ and $3 \times 3$ structures respectively.
As mentioned before, since the p2DEIA does not model the out-of-plane scattering, a small amount of energy will scatter away from the silicon slab in the 3D structure, hence a slight reduction in transmittance.
Therefore, to visually compare the transmittance in the 3D to those of the p2DEIA, the 3D values of $2 \times 2$ are multiplied by a factor of 1.2 and those of the $3 \times 3$ are multiplied by 1.5.
After applying the adjustment factors, the target, p2DEIA, and 3D values agree with each other.
To visually compare the measured transmittances to those of the target, similar scale factors of 1.45 and $1.2$ are needed for the $3 \times 3$ and the $2 \times 2$ respectively.
Therefore, the measured transmittances closely follow the values from the 3D model, indicating the fidelity of the model.
In conclusion, despite the constraints present in the p2DEIA model and the imperfection in the measurement, the good agreement between the target, p2DEIA, 3D, and  measured transmission values at $\lambda_0 = 1.525 \ \mu m$ is an indication of the success of the p2DEIA as a computational tool for inverse design of 3D planar structures.

In panels (b) and (d), the results of the wavelength sweep for the transmittances are illustrated. 
The solid thick curves are the simulation data from the 3D model and the thin solid lines are the measurement data. 
The agreement between the simulation and measurement data is excellent at the design wavelength (See SM for measurement details).
As mentioned before the magnitudes of the transmission parameters show a broadband response as one can see from the slow variation of the simulation and experimental data vs. wavelength.

\subsection*{Inverse Design of a silicon photonics metastructure for $10 \times 10$ vector-matrix Multiplication}
\begin{figure}[!ht]
    \centering
    \includegraphics[width = \textwidth]{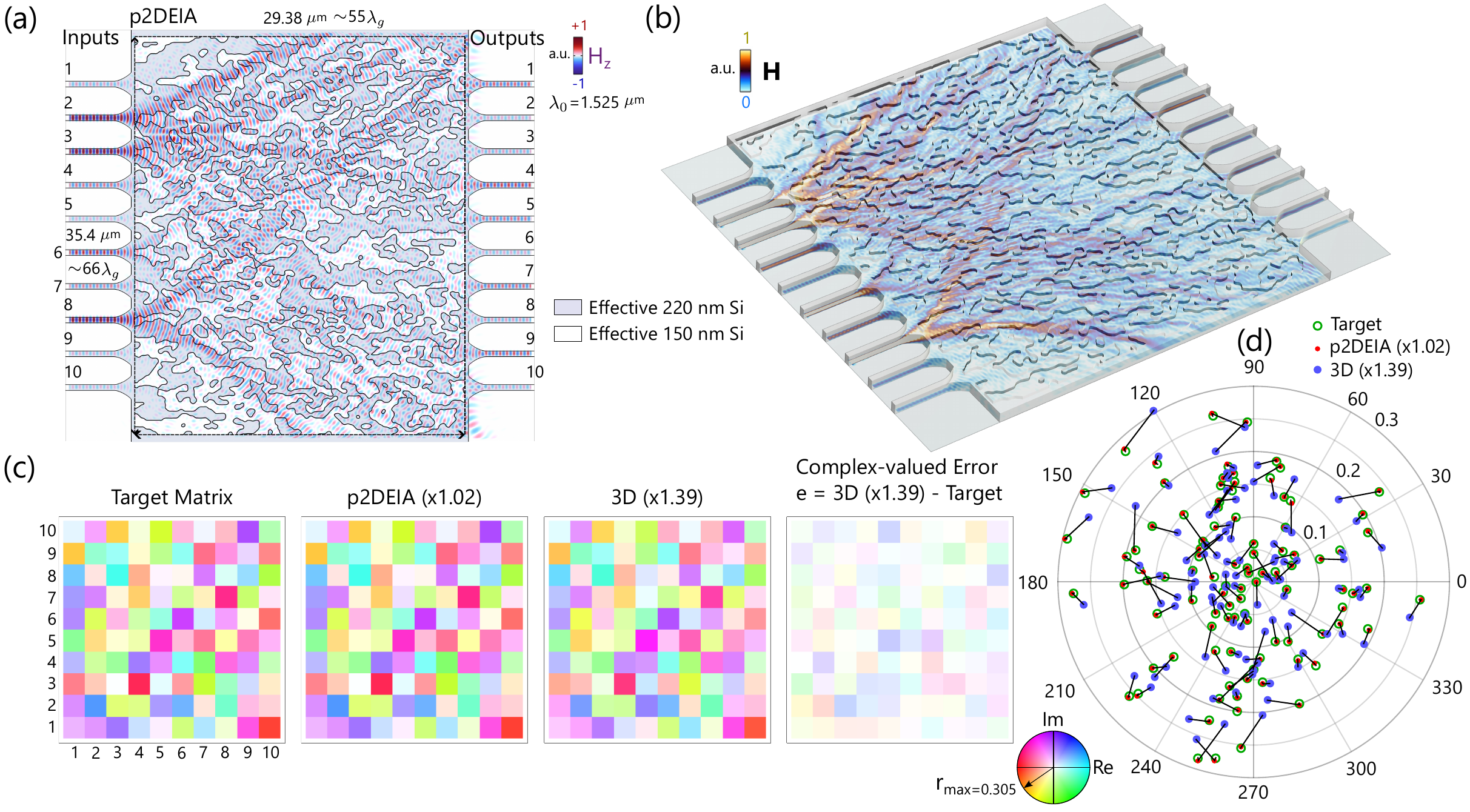}
    \caption{\textbf{Inverse design of a silicon photonics metastrutcure for performing a $10 \times 10$ vector-matrix multiplication} (a) The optimized geometry in p2DEIA and the distribution of the out-of-plane-polarized magnetic field for an arbitrary input excitation vector. (b) The absolute value of magnetic field distribution on a cut-plane through silicon in the 3D-rendered structure for the same excitation as in (a). (c) The target transmission matrix, the simulated transmission parameters in p2DEIA and 3D models, and the complex-valued error between the target and 3D transmission parameters. (d) The corresponding transmission values on complex plane polar plot.}
    \label{tenByten}
\end{figure}

As mentioned before, p2DEIA enables the simulation and design of an optically-large structure via tractable computational efforts.
This allows one to increase the number of the input and output data channels by incorporating more waveguides.
For that purpose, we used the same inverse-design technique, utilizing the p2DEIA to explore and design a structure to perform a $10 \times 10$ vector-matrix multiplication (see figure \ref{tenByten}(a)).
This was done within a $35.4 \ \mu m \times 29.4 \ \mu m$ design region using the same silicon photonics platform and wavelength as the previous examples.
The design region spans many guided wavelengths, i.e. $66\lambda_g \times 55 \lambda_g$ where $\lambda_g = \lambda_0/n_\text{eff, 220nm}$.
The choice of the target matrix $\mathbf{M}_{10 \times 10}$ was arbitrary within the space of passive transmission matrices (see the SM).

Figure \ref{tenByten}(a) illustrates the optimized geometry and a time snapshot of the out-of-plane magnetic field distribution for an arbitrary vector excitation in the p2DEIA model.
Figure \ref{tenByten}(b) shows the absolute value of the magnetic field distribution for the same excitation through the 3D-rendered structure.
The full-wave 3D simulation of such an exceedingly large optical structure is arduous using either frequency or time-domain simulations, however, with p2DEIA we were able to simulate the structure for hundreds of iterations during the inverse design procedure.
Figure \ref{tenByten}(c) illustrates the complex values of the target transmission matrix elements, the values from the optimized p2DEIA model, and the values from the 3D-rendered structure.
Similar to the $2 \times 2$ and $3 \times 3$ results, the p2DEIA, and 3D values are offset by proper adjustment factors.
For a more clear comparison, the corresponding transmission values are represented in the complex plane polar plot as shown in Figure \ref{tenByten}(d). (The proper adjustment factors are shown in parentheses.) 

The agreement between the target and p2DEIA values is excellent, signifying that the inverse design converged nicely. 
The 3D values show an acceptable agreement with the p2DEIA values despite the use of a coarser meshing and reduced order absorbing boundary conditions forced upon by the huge size of the 3D simulation.
However, the p2DEIA model introduces some errors on each interface (see SM).
Larger structures might require more sophisticated interface modeling that should take into account out-of-plane scattering and phase discontinuities due to the stored energy in non-propagating modes.
Regardless, the optimized design performs well judging by the close agreement between the target, the p2DEIA, and the 3D transmission parameters.
Therefore, by this design, p2DEIA proved to be successful in designing an optically-large structure for manipulating a large number of input and output data channels. 

\section*{Summary and Conclusion}
In this paper, we employed the p2DEIA using the effective index of the slab mode as an efficient computational platform for the inverse design of 3D planar structures in silicon photonics.
Assuming a low-index contrast, we demonstrated inverse-designed metastructures for performing $2 \times 2$ and $3 \times 3$ vector-matrix products.
The results of the simulation and experiment illustrate the high accuracy of the p2DIEA for approximating 3D planar structures.
Furthermore, the essential assumption of the low-index contrast for p2DEIA leads to feed-forward, low-resonant structures in which the response is slowly varying as a function of wavelength. 
Therefore, the resulting structures using the p2DEIA present broadband optical responses for performing analog computations and are also less sensitive to fabrication errors. 
Since the p2DEIA is a reduced-dimension approximation of a 3D structure, scaling to larger structures is much less of a computational burden than full-wave 3D simulation, thus enabling the inverse design of metastructures for larger-scale vector-matrix multiplications. 
We demonstrated this by designing a structure for performing a $10 \times 10$ vector-matrix product.

\section*{Acknowledgment}
This work is supported in part by the US Air Force Office of Scientific Research (AFOSR) Multidisciplinary University Research Initiative (MURI) grant number FA9550-21-1-0312 (to N.E.) and in part by the US Office of Naval Research (ONR) grant number N00014-19-1-2248 (to F.A.).

\bibliography{bibliography}

\bibliographystyle{Science}

\section*{Supplementary materials}
\subsection*{Effective-Index Approximation Overview}
The ever-increasing interest in wave-based analog computing demands efficient numerical methods for the simulation and design of optically-large electromagnetic structures.
One way to alleviate the computational effort is by approximating a structure with an effective model.
An effective model is meant to replace the original problem with a simpler one that is crafted to capture the essential features of the electromagnetic features in question.
The computational effort of the original problem might rapidly scale with the size of the structure, however, while the effective model of it might scale more favorably with regards to computational resources.
One way of formulating an effective model is through dimensionality reduction.
One example is the Eigenmode Expansion Method (EME) in which a waveguide of arbitrary cross-section (but uniform in $z$) is reduced to several guided modes, each propagating in 1D according to their index.
This removes two dimensions of computational effort while leaving one to be handled via appropriate means such as FDTD.
Furthermore, in the variational method for planar dielectric structures, one may replace the 3D structure with a two-dimensional effective index distribution for emulating the in-plane wave propagation in the actual 3D structure\cite{hammer2009effective}.
The effective index distribution is optimized by minimizing the functional form of Maxwell's equations for estimating the in-plane wave propagation.

In this work, the traditional effective model based on slab mode effective index (p2DEIA) is utilized for the inverse design of 3D planar structures based on silicon photonics.
In a typical silicon planar structure as shown in Figure \ref{fig:effectiveIndex}(a), the wave is confined within the silicon core in the form of a slab mode.
One can change the propagation characteristics of a slab mode, by changing the thickness of the core via etching down into some desired areas.
The resulting discontinuities between the etched and non-etched regions create interfaces through which the wave reflects and refracts and one can manipulate the light propagating from one side to the other by engineering the topology and shape of the interfaces.
The computational cost of full-wave 3D simulations of Maxwell's equations might render a design process highly inefficient.
Specifically, the computational cost grows in some methods such as the inverse design in which one needs to simulate the structure hundreds of times during the optimization process.
On the other hand, in silicon planar structures, the wave mostly propagates in the form of the fundamental slab mode in different regions (see Figure \ref{fig:effectiveIndex}(a)).
Therefore, the wave function along the out-of-plane direction ($y$-direction in Figure \ref{fig:effectiveIndex}(a)) is known provided the wave couples to a fundamental slab mode as it propagates through interfaces between etched and non-etched regions, that is, the coupling between the fundamental slab mode in each region and higher-order modes is negligible.
Therefore, for simulating the wave propagation efficiently throughout the planar structure in Figure \ref{fig:effectiveIndex}(a), one can simulate the \emph{in-plane} slab mode propagation (the $xz-plane$ in Figure \ref{fig:effectiveIndex}(a)) by an \emph{effective} two-dimensional model.
In the effective 2D model as illustrated in Figure \ref{fig:effectiveIndex}(b), the regions and interfaces of the structure in Figure \ref{fig:effectiveIndex}(a) are mapped onto a two-dimensional plane.
Therefore, a 2D scalar wave satisfying Helmholtz's wave equation represents the in-plane propagation of the slab mode throughout the structure in Figure \ref{fig:effectiveIndex}(a).
\renewcommand{\thefigure}{S1}
\begin{figure}[ht!]
    \centering
    \includegraphics[width = \textwidth]{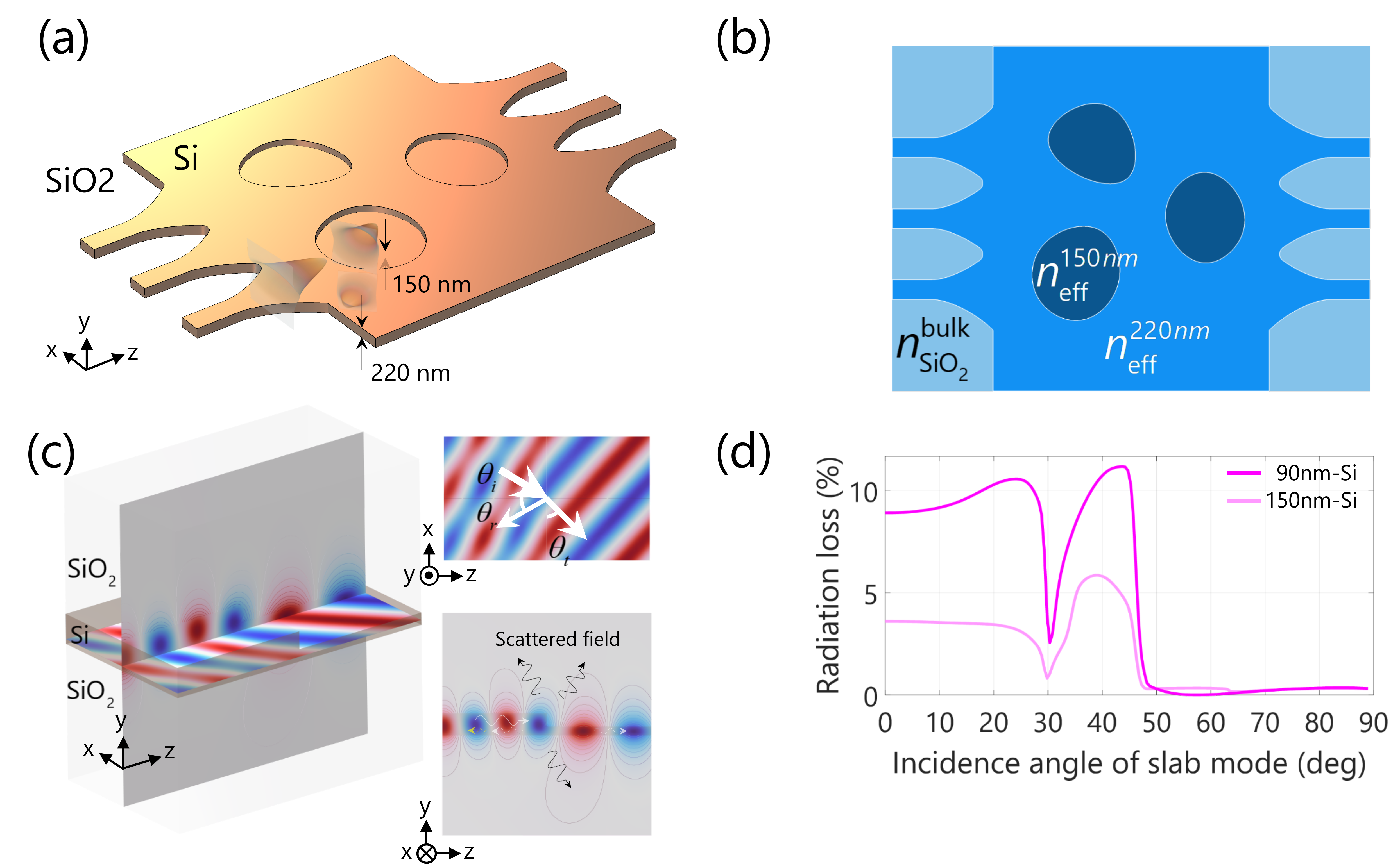}
    \caption{\textbf{Effective index approximation for planar structures} (a) The 3D model of a typical silicon planar structure with etched regions for manipulating the in-plane wave propagation (b) Effective index approximation method for projecting the 3D model onto an effective 2D model (p2DEIA) for capturing the main features of the in-plane wave propagation (c) The full-wave 3D simulation of a slab mode propagation impinging on an etched interface with incidence angle $\theta_i$. On the $ xz$ plane, the primary in-plane wave propagation, which is captured by p2DEIA, is illustrated. On the $yz$ plane the secondary out-of-plane scattering due to height discontinuity is shown that can not be captured by p2DEIA (d) The energy of the out-of-plane scattering (radiation loss) vs. the angle of incidence, $\theta_i$.}
    \label{fig:effectiveIndex}
\end{figure}
The refractive index of the regions in the 2D model is assigned from the effective indices of the slab modes confined in the respective regions in the 3D structure.
As a result of this modeling, the phase progression of the slab mode through the in-plane geometry is accurately modeled as long as the slab mode propagates without encountering an interface.
In other words, the bulk phase progression can be accurately modeled using the p2DEIA.

Focusing on phase progression comes at a cost with regard to accuracy when an interface is encountered.
The ground truth for interface behavior is the full 3D model as illustrated in Figure. \ref{fig:effectiveIndex}(c).
The wave propagation through the 3D interface exhibit a few effects not included in the 2D model such as the out-of-plane scattering as shown by the field distribution on the $yz$ plane in Figure \ref{fig:effectiveIndex}(c).
In contrast, simulating plane wave illumination of a planar interface between two materials of differing effective indices using the p2DEIA model will result in the standard Fresnel reflection and transmission coefficients.
As a meaningful way to understand the interface errors introduced by the p2DEIA model, we can compare the Fresnel reflection and transmission coefficients to the reflection and transmission of the fundamental slab modes in the full 3D model as a function of the angle of incidence.

The channeling to the secondary modes (out-of-plane scattering), which we call the radiation loss, causes the reflection and transmission of the fundamental slab modes to slightly deviate from the Fresnel reflection and transmission coefficients calculated using the 2D model.
In Figure \ref{fig:effectiveIndex}(d), we plot the fraction of the input power that channels to radiation loss i.e $L = 1-|R|^2-|T|^2$ vs. the angle of incidence for the $220 \ nm$ to $150 \ nm$ and the $220 \ nm$ to $90 \ nm$ interfaces.

As one can see, the amount of radiation loss scales directly with the height of the etch while following similar angular dispersion. 
Also, it peaks around the critical angle of incidence beyond which total internal reflection happens and the radiation loss becomes smaller.
Therefore, the accuracy of the p2DEIA relates to the etch height, that is, the larger the etch height the more inaccurate the p2DEIA is.
In our design, we try to control the error of the p2DEIA by using a relatively shallow etch of $70 \ nm$ (corresponding to a $150 \ nm$-thick silicon slab).
Note that while $150 \ nm$ presents a maximum radiation loss $\approx\%6$ just prior to the critical angle, the average radiation loss is lower when considering all possible angles.

The amplitude and phase of the transmission coefficients through interfaces with different etch depths are plotted in Figure \ref{fig:transmission}.
\renewcommand{\thefigure}{S2}
\begin{figure}[ht!]
    \centering
    \includegraphics[width = \textwidth]{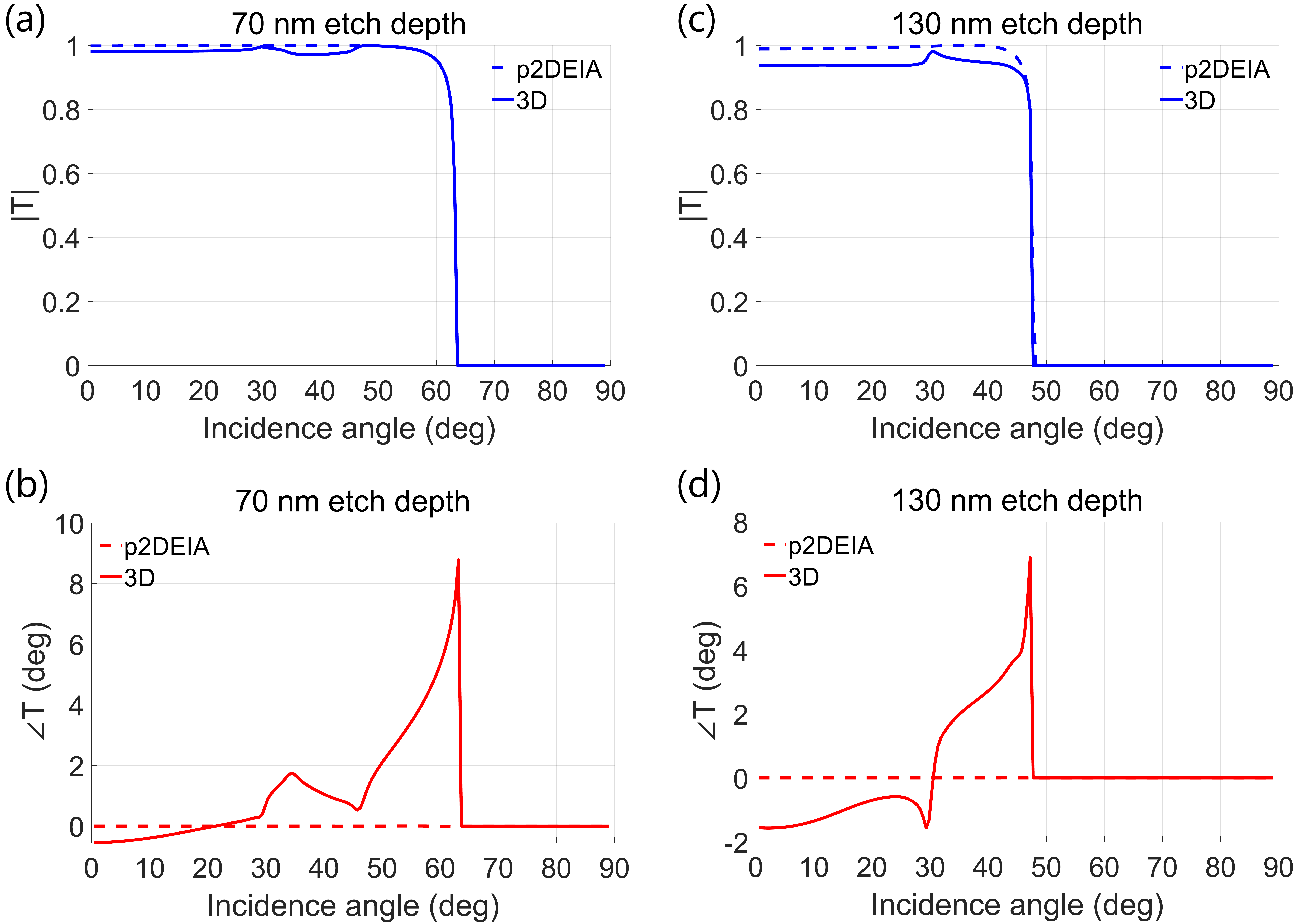}
    \caption{\textbf{Interface transmission coefficient} (a,b) The magnitude and phase of the transmission coefficient for the 3D 220nm to 150nm-thick silicon interface (the solid curves) and for the corresponding projected interface in the p2DEIA model (dashed curves). (c,d) The corresponding plots for the 220nm to 90nm silicon interface}
    \label{fig:transmission}
\end{figure}
As one can see, beyond the critical angle the data for 3D and p2DEIA coincide since there is negligible radiation loss due to total internal reflection.
In the p2DEIA model, the effective index contrast is greater for the 130nm etch than the 70nm etch and naturally, this will reduce the transmission magnitude.
However, this difference is slight and transmission is near unity for both.
Before the critical angle, one can observe a far greater discrepancy in the 130nm etch compared to the 70nm etch between the 3D model and p2DEIA due to the out-of-plane scattering modes.
The reason for this transmission drop is poorer mode overlap (both shape and mode center) between the fundamental modes yielding increased reflection, 3D radiation loss, and scattering into higher-order guided modes (see Figure \ref{fig:reflection} for comparing the reflections for p2DEIA and 3D).
Also, from the phase plots in Figure \ref{fig:transmission}(b) and (d), one can observe that in the region near the critical angle the phase discrepancy between the 3D model and the p2DEIA spikes due to the heightened mismatch. 
However, before the critical angle, the phase error profile is smaller for $70 \ nm$-etch height than the $130 \ nm$ case which further justifies the use of $70 \ nm$-etch height for better accuracy.

For completeness, we also present analogous data for the reflection coefficients in Figure \ref{fig:reflection}.
As a generally feed forward system, this is of less importance, however it is clear that the reflection magnitude is less for the shallower etch depth.

In a practical design, the wave will interact with many interfaces between the input and output.
Each interaction will introduce some power loss and a slight phase rotation the accumulation of which results in an overall loss factor and phase error.
\renewcommand{\thefigure}{S3}
\begin{figure}[ht!]
    \centering
    \includegraphics[width = \textwidth]{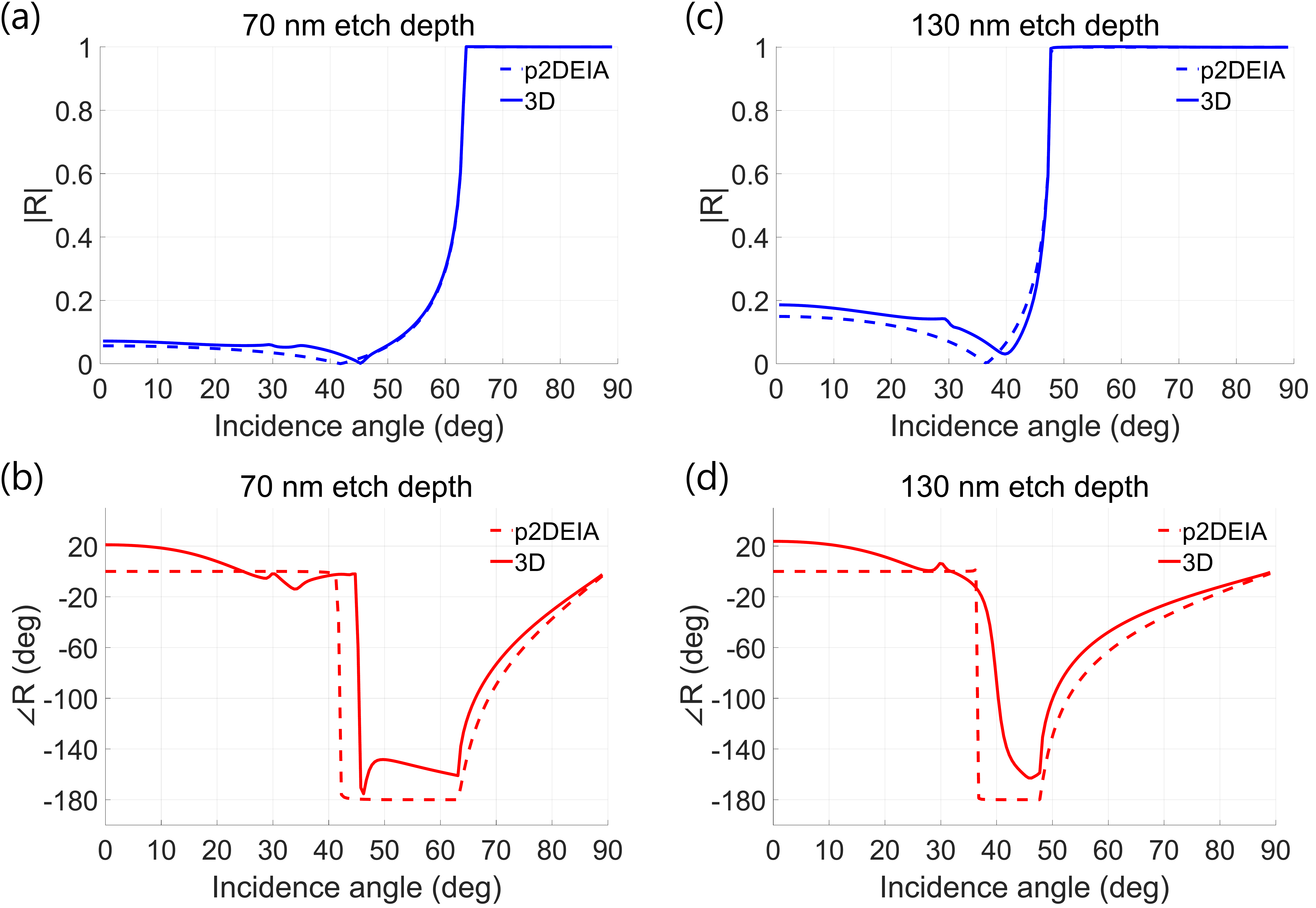}
    \caption{\textbf{Interface reflection coefficient} (a,b) The magnitude and phase of the reflection coefficient for the 3D 220nm to 150nm-thick silicon interface (the solid curves) and for the corresponding projected interface in the p2DEIA model (dashed curves). (c,d) The corresponding plots for the 220nm to 90nm silicon interface}
    \label{fig:reflection}
\end{figure}
\subsection*{Normalizing the scattering error}

We desire to consider a meaningful relative error for our devices.
Consider an error defined on a realized experimental $N \times N$ matrix $\mathbf{T}_\mathrm{exp}$ when compared to a similar target matrix $\mathbf{T}_\mathrm{targ}$.
A natural metric for the squared error is the square of the Frobenius norm, $\lVert \mathbf{T}_\mathrm{exp} -  \mathbf{T}_\mathrm{targ} \rVert_F^2$.
However, this metric makes it difficult to compare matrices of different sizes as a larger matrix will naturally have a larger norm than a smaller one of similar values.
On the other hand, these matrices will likely \emph{not} have similar values as the elements of a passive scattering matrix will naturally decrease with the size of the matrix.
This potentially will make the error appear smaller for larger matrices.  
We seek an error metric that allows us to compare devices in spite of these complications.

Let us consider the \emph{average relative squared error} of the matrix terms $T_{i,j}$.
The average relative squared error is $\lVert \mathbf{T}_\mathrm{exp} -  \mathbf{T}_\mathrm{targ} \rVert_F^2/(N^2 \Delta_\mathrm{typ}^2)$ where $\Delta_\mathrm{typ}$ is some typical complex distance value and $N^2$ is the number of terms in matrix.
In many instances in physics and engineering, the typical distance value is taken to be the \emph{target} value itself.
However, such a scheme over-emphasizes the importance of small target values.
Indeed, if a single target term would be zero, any experimental realization would have infinite relative error even if practically it performed quite well.
Rather, we note that we are interested in target matrices that are unitary (or close to it) and will have elements with magnitudes on the order of $|\mathbf{T}_{i,j}| \approx 1/\sqrt{N}$.
Since we are interested in the \emph{complex} error, a realized experimental value that is 180 degrees out of phase would possess a typical absolute error distance,  $\Delta_\mathrm{typ}$,  of $2/\sqrt{N}$.
Therefore, the average relative squared error becomes $\mathrm{error}^2 = \lVert \mathbf{T}_\mathrm{exp} -  \mathbf{T}_\mathrm{targ} \rVert_F^2/(4 N)$.

This metric has the following useful properties.  Suppose that the target matrix, $\mathbf{T}_\mathrm{targ}$, is unitary.
If $\mathbf{T}_\mathrm{exp}$ = -$\mathbf{T}_\mathrm{targ}$, then $\mathrm{error}^2 = 1.00$.
If $\mathbf{T}_\mathrm{exp}$ = $\mathbf{0}$, then $\mathrm{error}^2 = 0.25$.
If $\mathbf{T}_\mathrm{exp}$ = $\mathbf{T}_\mathrm{targ}$, then $\mathrm{error}^2 = 0.00$ regardless of any zero terms.

\subsection*{Fabrication}
\renewcommand{\thefigure}{S4}
\begin{figure}[ht!]
    \centering
    \includegraphics[width = \textwidth]{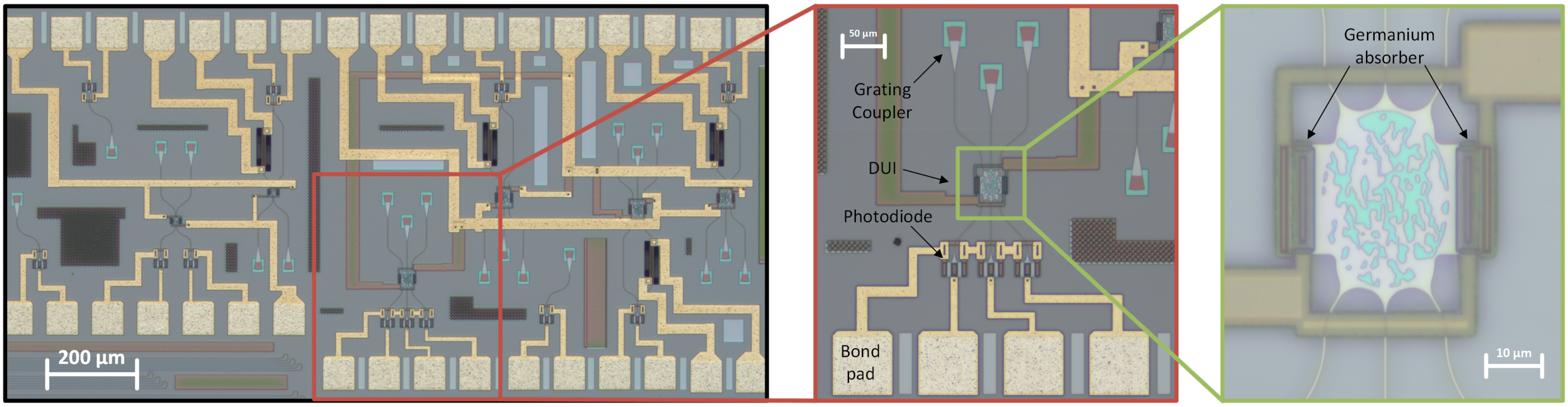}
    \caption{
        Micrograph of the physical experiment showing several $2 \times 2$ and $3 \times 3$ kernels and various calibration structures.  The $3 \times 3$ kernel is highlighted.
    }
    \label{fig:chip_photo}
\end{figure}
The photonic chip was fabricated in the AMF \SI{180}{\nano\metre} SOI process with a \SI{2}{\micro\metre} thick buried oxide. 
The photonic signal is routed on chip using single mode waveguides with \SI{500}{\nano\metre} width and \SI{220}{\nano\metre} height.
The light is coupled to the chip using grating couplers with efficiency of about \SI{40}{\%}. 
The AMF process also provides photodiodes with responsivity of about \SI{0.8}{\A/\W} that are used to monitor the output optical power.

\subsection*{Measurement}
\renewcommand{\thefigure}{S5}
\begin{figure}[ht!]
    \centering
    \includegraphics[width = \textwidth]{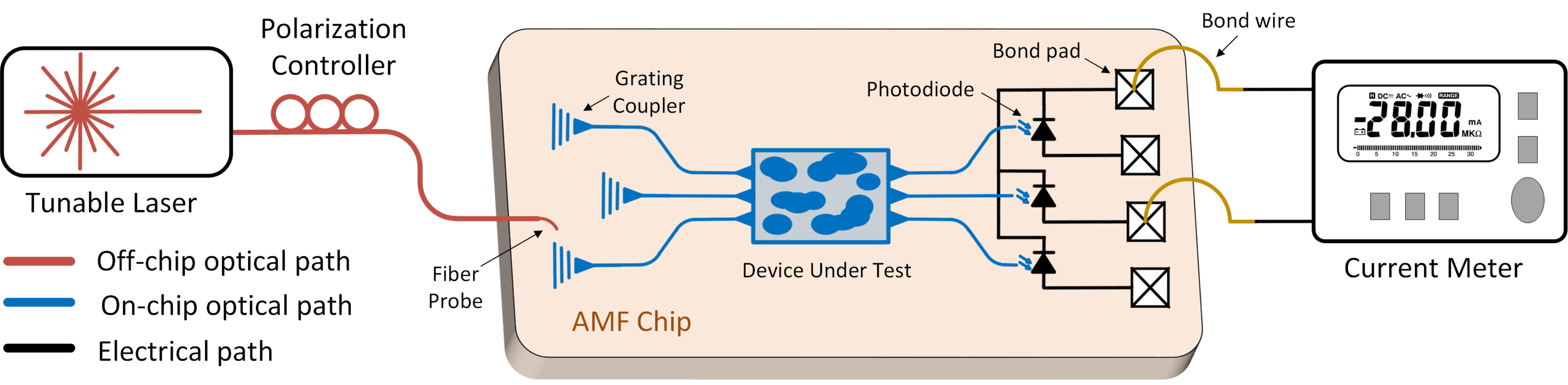}
    \caption{
        Measurement setup.  A tunable laser is used to illuminate one of several grating couplers.  The light moves through the inverse-designed structure. Finally, the power is measured at each of the three photodiodes with a current meter.}
    \label{fig:Measurement Setup}
\end{figure}
A tunable laser is used to test the performance of the fabricated inverse-design devices. 
As shown in Figure \ref{fig:Measurement Setup}, the infrared light is launched to the on-chip grating coupler using a fiber probe. 
The light polarization is adjusted using a polarization controller to match the grating coupler’s polarization response and maximize the coupled optical power to the on-chip waveguide. 
At the output, photodiodes are implemented to convert the optical power to an electrical current which later is measured by a current meter.

\subsection*{Analyzing the computational resources for 3D and p2DEIA structures}

A natural question is what is quantitatively gained by performing the inverse-design method using the p2DEIA vs the full-wave 3D model.
Since the inverse design may not be computationally feasible for large structures, we will employ a heuristic analysis.
The inverse design utilizes the Adjoint method in which the gradient is calculated from a series of simulations in which both the ``input'' and ``output'' ports are illuminated.
Given a chosen scattering parameter objective, whether the p2DEIA or 3D model is being optimized, the same number of simulations per optimization iteration is required.
Therefore, to first order, in order to compare the computational effort required to perform an optimization, we need only compare the costs of solving for the optical fields in both cases.
However, this is only approximate, as are many details which can affect the convergence in both cases.

In this section, we compare the computational resources needed to solve for the optical fields in the full-wave 3D and p2DEIA models.
As for the case in the study, we consider the silicon photonics metastructure designed to perform $10 \times 10$ vector-matrix product (see the main text).
We examine the structure by considering one random input vector of modes over the input ports.
We employ the finite element method (FEM) to simulate time-harmonic optical fields at the design wavelength $\lambda_0 = 1.525 \ \mu m$ using an Intel(R) Xenon(R) Gold 6130 CPU @ 2.10GHz with 16 cores and 225GB of RAM.
For solving the linear systems, we employ the Generalized Minimum Residual Method (GMRES) iterative solver\cite{saad1986gmres} and parallel direct solver PARDISO\cite{schenk2002two} using COMSOL 5.6.
\renewcommand{\thefigure}{S6}
\begin{figure}[ht!]
    \centering
    \includegraphics[width = \textwidth]{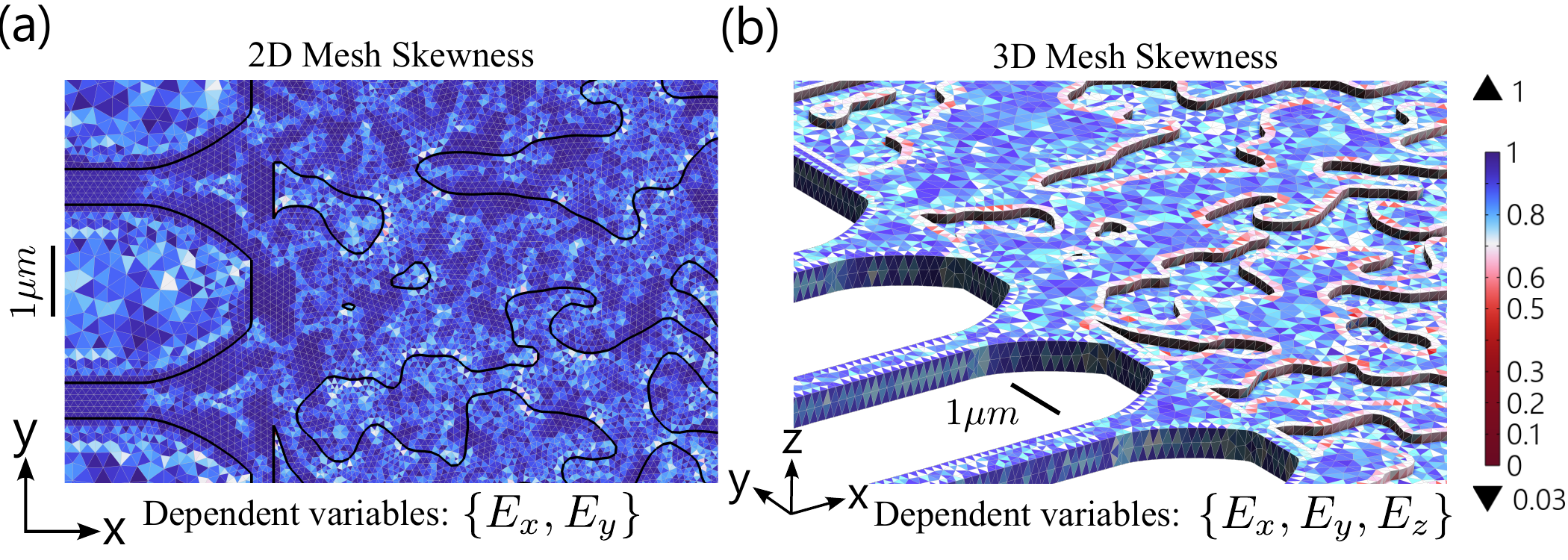}
    \caption{\textbf{Mesh skewness} (a) Mesh plot for the p2DEIA model of the $10 \times 10$ silicon photonics metastructure (b) The corresponding plot for the 3D model}
    \label{fig:Mesh-Skewness}
\end{figure}
Table 1 includes information about the domain discretization, resources, and time needed to solve the linear systems as detailed below:
\begin{itemize}
  \item \textbf{Mesh volume(area):} The volume(area) of the simulation domain of the 3D(p2DEIA) structure.
  \item \textbf{Mesh cells (order)}: The total number of cells discretizing the simulation domain. For the 3D simulation, the majority of the cells are tetrahedral. For the p2DEIA simulation, the cells are triangular. Here \emph{order} refers to the order of the shape function associated with individual cells.
  \item \textbf{Average quality}: We considered the mesh skewness as the quality factor as shown in Fig.\ref{fig:Mesh-Skewness}. It quantifies how deformed a cell is compared with a reference cell that is optimal for discretization. The skewness merit varies between 0 (bad) and 1 (good). The average of the skewness merit over the simulation domain is reported.
  \item \textbf{Degrees of freedom (DOF)}: The number of unknowns that characterize the optical fields. It depends on the number of electric field components considered as the dependent variables, the shape function, and the number of cells. According to the case in the study, for the p2DEIA simulation, we consider $\{E_x,E_y\}$ and for the 3D model we consider all three components, $\{E_x,E_y,E_z\}$.
  \item \textbf{Physical memory (RAM)}: The size of the physical memory that is occupied by the linear systems.
  \item \textbf{Time}: The amount of time that is needed to solve the linear systems.
\end{itemize}

\begin{tabular}{ |p{2.4cm}|p{1.8cm}|p{1.5cm}|p{1.5cm}|p{1.5cm}|p{1.5cm}|p{1.8cm}| }
 \hline
 \multicolumn{7}{|c|}{Computational Resources Analysis Per One Random Excitation (FEM method)} \\
 \hline
 EM model \break (Solver)&Mesh Volume \break (Area)& cells \break (order)&Average \break Quality&DOF&RAM& Time\\
 \hline
 p2DEIA \break (PARDISO)&788 ${\mu \text{m}}^2$&181K \break (Cubic)  &0.85&1.92M&8.01GB&4.9sec\\
 \hline
 p2DEIA \break (GMRES)&788${\mu \text{m}}^2$&181K \break  (Cubic) &0.85&1.92M&3.4GB&16min\\
 \hline
 3D \break (GMRES)&3058${\mu \text{m}}^3$&926K \break (Cubic)&0.63&18.42M&186GB&5h 45min\\
 \hline
 3D \break (PARDISO)&3058${\mu \text{m}}^3$&926K \break (Cubic)&0.63&18.42M&400GB&N.A.\\
 \hline
\end{tabular} \\

The data shows the p2DEIA model can significantly speed up the inverse design process by fast evaluation of the optical field in each iteration while using fewer computational resources.

\end{document}